\documentclass[11pt,a4paper]{amsart}

\usepackage[latin1]{inputenc}
\usepackage[english]{babel}
\usepackage{amsmath}

\usepackage{amssymb, mathabx}
\usepackage{graphicx}
\usepackage{braket}
\usepackage[pdftex,plainpages=false,colorlinks,hyperindex,bookmarksopen,linkcolor=red,citecolor=blue,urlcolor=blue]{hyperref}

\usepackage{cite}

\usepackage{mathrsfs}

\usepackage{epstopdf}

\usepackage{braket}

\usepackage[hmargin=3cm,vmargin={3.5cm,4cm}]{geometry}

\theoremstyle{theorem}

\theoremstyle{definition}                                 

\theoremstyle{definition}                           
\newtheorem{defin}{Definition}                   
\newtheorem{example}{Example}

\theoremstyle{remark}                             
\newtheorem*{rmk}{Remark}              

\usepackage{color}

\usepackage{mathtools,slashed}

\newcommand{\be}{\begin{eqnarray}}
\newcommand{\ee}{\end{eqnarray}}

\newcommand{\R}{\mathbb{R}}  
\def\eg{{\it e.g. }} 
\def\ie{{\it i.e. }}

\newcommand{\ceil}[1]{\lceil #1 \rceil}

\def\eg{{\it e.g.}\ }
\def\ie{{\it i.e.}\ }

\numberwithin{equation}{section}

\allowdisplaybreaks

\begin{document}
\title{Scott-Blair models with time-varying viscosity}
	
	   \author{Ivano Colombaro$^1$}
		\address{${}^1$ Department of Information and Communication Technologies, Universitat Pompeu Fabra and INFN.
		C/Roc Boronat 138, Barcelona, SPAIN.}
		\email{ivano.colombaro@upf.edu}
	
		 \author{Roberto Garra$^2$}
		\address{${}^2$ University of Rome La Sapienza, Dipartimento di Scienze Statistiche,
		P. le A. Moro 5, 00185 Roma, ITALY   
    	    }
    	    \email{roberto.garra@sbai.uniroma1.it}
	
	    \author{Andrea Giusti$^3$}
		\address{${}^3$ Department of Physics $\&$ Astronomy, University of 	
    	    Bologna and INFN. Via Irnerio 46, Bologna, ITALY and 
	    	 Arnold Sommerfeld Center, Ludwig-Maximilians-Universit\"at, 
	    	 Theresienstra{\ss}e~37, 80333 M\"unchen, GERMANY.   
    	    }
		\email{agiusti@bo.infn.it}
	
    \author{Francesco Mainardi$^4$}
    	    \address{${}^4$ Department of Physics $\&$ Astronomy, University of 	
    	    Bologna and INFN. Via Irnerio 46, Bologna, ITALY.}
			\email{francesco.mainardi@bo.infn.it} 
 

	\thanks{In: \textbf{Applied~Mathematics~Letters~86~(2018) pp.~57--63 
}, \textbf{DOI}: \href{https://www.sciencedirect.com/science/article/pii/S089396591830199X}{10.1016/j.aml.2018.06.022}}	
	
    \date  {\today}


\begin{abstract}
In a recent paper, Zhou et al. \cite{zhou} studied the time-dependent properties of Glass Fiber Reinforced Polymers composites by employing a new rheological model with a time-dependent viscosity coefficient. This rheological model is essentially based on a generalized Scott-Blair body with a time-dependent viscosity coefficient. Motivated by this study, in this note we suggest a different generalization of the Scott-Blair model based on the application of Caputo-type fractional derivatives of a function with respect to another function. This new mathematical approach can be useful in viscoelasticity and diffusion processes in order to model systems with time-dependent features. In this paper we also provide the general solution of the creep experiment for our improved Scott-Blair model together with some explicit examples and illuminating plots.
\end{abstract}

\maketitle

\section{Introduction}

	In the 50's, the British chemist George William Scott-Blair introduced the first linear viscoelasticty model with physical properties intermediate between the elastic Hooke model and the fluid Newton model. The mathematical implementation of Scott-Blair's ideal system was performed by means of the notion of derivative of non-integer order, see \eg \cite{Rogosin-Mainardi_CAIM2014} for an historical perspective.
	
        Unfortunately, as reported by Stiassnie in \cite{Stiassnie_AMM1979}, after several experimental contributions in the emerging new field of viscoelasticity, ultimately Scott-Blair gave up on studying the implications of non-integer order calculus in rheology as he could not find a satisfactory mathematical definition of ``fractional differential''.

	Since the time of Scott-Blair, the theory of fractional calculus \cite{Caputo-Mainardi_RNC1971, Mainardi-Tomirotti_GEO1997, Mainardi_BOOK2010, Mainardi-Spada_EPJ-ST2011, Giusti-NODY}, that is the mathematical discipline dealing with integrals and derivatives of non integer order, has been refined and represents one of the most important languages for the modern formulation of linear viscoelasticity \cite{Mainardi_BOOK2010, Mainardi-Spada_EPJ-ST2011}. Besides, it is worth noting that fractional calculus plays a central role also in many other fields of science, see \eg \cite{ERT, Garrappa-CNSNS, Sandev, G-JMP} and references therein.
	
	In a recent paper, Zhou et al. \cite{zhou} investigated the time-dependent properties of Glass Fiber Reinforced Polymers (GFRP) composites by means of a generalized Scott-Blair model with time-varying viscosity. Furthermore, it is worth remarking that viscoelastic models featuring a time-varying viscosity have also been analyzed in \cite{Pandey-Holm_PRE2016} by Pandey and Holm. 
 
	Motivated by these studies, in this paper we perform a generalization of the Scott-Blair model by employing the so called Caputo fractional derivatives of a function with respect to a given function. In our view, this approach can be particularly useful to especially introduce the time dependency of diffusion or viscosity coefficients right at the fractional level. For example, a new fractional Dodson diffusion model based on this approach was studied in \cite{noi}. 
	
	The general interest for the potential applications of this new approach is supported by some recent studies, see \eg \cite{almeida,alm1,alm2,orega}. Hence, the aim of this explorative study is to suggest potential applications of the Caputo fractional derivative of a function with respect to a given function in rheology and material science.
	

\section{Mathematical preliminaries} \label{preliminaries}
	Fractional derivatives of a function with respect to another function have been known, though soon forgotten, objects since the classical monograph by Kilbas et al. \cite{kilbas} (Section 2.5). Nonetheless, thanks to a recent paper Almeida \cite{almeida}, the Caputo-type regularization of these fractional operators has undergone a rebirth in recent years.
	
	\begin{defin}
	Let $\nu>0$, $I = (a,b)$ be an interval such that $-\infty \leq a < b \leq +\infty$, $g(t) \in L_1 (I)$ and $f(t)\in C^1(I)$ strictly increasing function for all $t\in I$.
	Then, the fractional integral of a function $g(t)$ with respect to another function $f(t)$ is given by 
\begin{equation}
I^{\nu,f}_{a+}g(t):=\frac{1}{\Gamma(\nu)}\int_a^t f'(\tau)
[f(t)-f(\tau)]^{\nu-1}g(\tau)d\tau.
\end{equation}
\end{defin}
 
\begin{rmk}
For $f(t) = t^{\alpha/\beta}$ we recover the definition of Erd\'elyi-Kober fractional integral that has recently found many applications in various branches of physics and mathematics, see \eg \cite{gia1, noi2}. Moreover, if we set $f(t)= \ln t$ we get the Hadamard fractional integral, whereas for $f(t)= t$, the Riemann-Liouville fractional integral (see \cite{kilbas}).
\end{rmk}
          
          The corresponding Caputo-type evolution operator is then given by
          \begin{defin} \label{def-der}
        Let $\nu>0$, $n \equiv \ceil{\nu}$, $I = (a,b)$ be an interval such that $-\infty \leq a < b \leq +\infty$, $g(t) \in C^n (I)$ and $f(t)\in C^1(I)$ strictly increasing function for all $t\in I$. 
        Then, the Caputo derivative of the function $g(t)$ with respect to a function $f(t)$ is given by
           \begin{equation}\label{2.1}
           {}^C \left(\frac{1}{f'(t)}\frac{d}{dt}\right)^\nu g(t) :=I_{a+}^{n-\nu,f}\left(\frac{1}{f'(t)}\frac{d}{dt}\right)^n g(t) \, .
               \end{equation} 
          \end{defin}

	Notice that, we used a quite different notation with respect to the one adopted by Almeida in \cite{almeida}. This choice was made in order to explicitly stress that one can, somehow, understand these operators as the fractional power counterpart of a sort of stretched derivative, \ie $\frac{1}{f'(t)}\left(\frac{\partial}{\partial t}\right)$.
          
          A key feature of the operator \eqref{2.1} is that if $g(t) =[f(t)-f(a+)]^{\beta-1}$ with $\beta>1$, then (see Lemma 1 of \cite{almeida})
          \begin{equation}\label{2.2}
        {}^C \left(\frac{1}{f'(t)}\frac{d}{dt}\right)^\nu  g(t) = \frac{\Gamma(\beta)}{\Gamma(\beta-\nu)}
        [f(t)-f(a+)]^{\beta-\nu-1}.
          \end{equation}
          
          As a consequence, the composition of the Mittag-Leffler function with the function $f(t)$, namely
          $g(t) = E_\nu[\lambda(f(t)-f(a+))^\nu]$, is an eigenfunction of the operator \eqref{2.1} with eigenvalue $\lambda$ (see \cite{noi}).
          
          For further details on these operators, we invite the interested reader to refer to \cite{almeida, noi, kilbas}.
 
 
      \section{A class of Scott-Blair models with time-varying viscosity}
	In a recent paper Zhou et al. \cite{zhou} studied the time-dependent property of GFRP composites, considering a time-dependent deformation at various stress level. This model is based on a classical Scott-Blair dashpot with a time dependent viscosity coefficient. Specifically, the constitutive equation for the model in \cite{zhou} reads
	\be \label{1}
	\sigma(t) = \eta (t) \, \tau ^{\gamma-1}D_{0+} ^\gamma \epsilon(t), \quad \gamma\in [0,1], \, \quad t > 0 \, ,
	\ee	
	where $\sigma(t)$ and $\epsilon(t)$ denote the stress and strain functions, $\tau$ is the (constant) relaxation time of the system, $\eta(t)$ the time dependent viscosity coefficient and $D_{0+} ^\gamma$ the Riemann-Liouville derivative of order $\gamma$.
	
\begin{figure}[h!]
	\centering
	\includegraphics[scale=0.35]{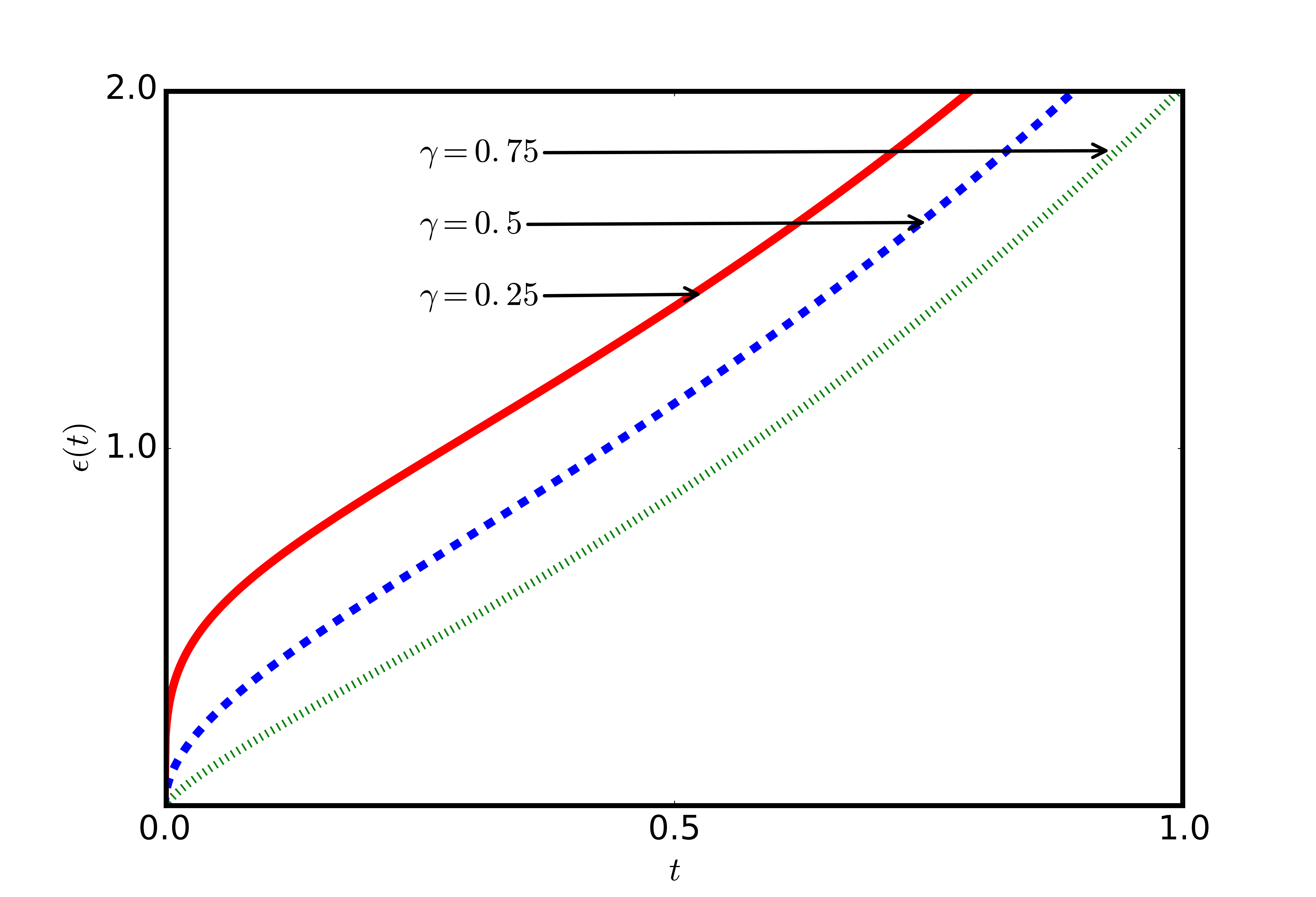}
	\caption{Plot of the strain response to the creep experiment in \eqref{3.3}, setting $\sigma _0$, $\tau$, $\eta _0$ and $\alpha$ to one.}
	 \label{fig-1}
	\end{figure}

	According to \cite{zhou}, if we now plug into \eqref{1} the condition
	\begin{equation}\label{2}
    \eta(t) = e^{- \alpha \, t} \, \eta_0 \ ,
    \end{equation}
    where $\alpha$ is an empirical parameter associated to the damage evolution, then one can easily find the solution of the creep experiment, \ie the solution of the constitutive equation \eqref{1} with the condition \eqref{2} and assuming a constant stress $\sigma (t) = \sigma _0$, is given by (see \cite{zhou})
	\begin{equation} \label{3.3}
    \epsilon(t)= \frac{\sigma_0 \, \tau}{\eta_0} \left( \frac{t}{\tau}\right)^\gamma \, E_{1,1+\gamma}(\alpha t),
    \end{equation}
    see \figurename~\ref{fig-1}, where $E_{1, 1+\gamma}(t) = \sum_{k=0}^{\infty}\left[ t^k / \Gamma(k+1+\gamma) \right]$, is a two-parameter Mittag-Leffler function.     
     
     We recall that this classical rheological model interpolates the pure solid behaviour for $\gamma = 0$ and the Newtonian fluid for $\gamma = 1$.
    
	From an experimental perspective, it is worth noting that this specific implementation of the Scott-Blair model with time-varying viscosity has been used, by the same authors, to characterize the damage growth during creep tests for salt rock samples, see \cite{zhou2}.

	Now, our proposal is to use a different mathematical approach based on the technique discussed in Section~\ref{preliminaries}. Indeed, let us consider a constitutive equation given by
	 \begin{equation}\label{3}
     \sigma(t) = \eta_0 \, \tau ^{\gamma-1} \, \mathcal{D}^\gamma \epsilon(t) \, , 
     \quad \mathcal{D}^\gamma \equiv {}^C \left(\frac{1}{f'(t)}\frac{d}{dt}\right)^\gamma \, ,
     \end{equation}
	with $\gamma \in [0, 1]$ and for some $f(t)\in C^1(\R^+)$ strictly increasing function for all $t>0$, \ie $\mathcal{D}^\gamma$ is defined according to \eqref{2.1} with $a = 0$ and $b = +\infty$.
	
	Thus, here the idea is to include the effect of the time dependence of the viscosity coefficient in the definition of the fractional operator.
     
    Now, using the results of Section~\ref{preliminaries}, one can infer the general solution of the creep experiment for a material described by the constitutive equation \eqref{3}. Indeed, once set $\sigma (t) = \sigma _0$, the general solution of \eqref{3} reads
    \be \label{4}
    \epsilon (t) = \frac{1}{\Gamma (\gamma + 1)} \frac{\sigma _0}{\eta _0 \, \tau ^{\gamma -1}} [f(t) - f(0+)]^{\gamma + 1} \, .
    \ee
    Indeed, considering the equation
    $$ \sigma _0 = \eta_0 \, \tau ^{\gamma-1} \, \mathcal{D}^\gamma \epsilon(t) \, , $$
    with the ansatz $\epsilon (t) = \mathcal{N} \, \big[ f(t) - f(0+) \big] ^{\beta - 1}$, with $\mathcal{N}$ being a normalization constant, using \eqref{2.2} one finds
    $$ \sigma _0 = \eta_0 \, \tau ^{\gamma-1} \, \frac{\Gamma(\beta)}{\Gamma(\beta-\gamma)}
        [f(t)-f(0+)]^{\beta-\gamma-1} \, . $$ 
Since the left-hand side of the latter is time independent, we are forced to set $\beta = \gamma + 1$. Thus, solving for $\mathcal{N}$ in the remaining equation one finds $\mathcal{N} = \sigma _0 / \eta _0 \, \tau^{\gamma - 1} \, \Gamma (\gamma +1)$, that confirms the result in \eqref{4}.

\begin{example} [Exponential function]
Let us consider the case with $f'(t) = e^{\alpha \, t}$. Then, \eqref{3} reads
\be 
 \sigma(t) = \eta_0 \, \tau ^{\gamma-1} \, \left(e^{- \alpha \, t} \, \frac{d}{dt}\right)^\gamma \epsilon(t) \, ,  
\ee
where we omitted the ${}^C$ for the sake of clarity.

Then, it is easy to see that $f(t) = e^{\alpha \, t} / \alpha + C$ with $C$ being an integration constant. Now, from \eqref{4} one immediately infers that the general solution for the latter equation is given by (see \figurename~\ref{fig-2})
\begin{equation} \label{3.8}
  \epsilon(t) = \frac{1}{\Gamma(\gamma+1)}\frac{\sigma_0}{\eta_0 \tau ^{\gamma-1}}\left(\frac{e^{\alpha t}-1}{\alpha}\right)^{\gamma +1}.
\end{equation}
It is worth noting that, again, for $\gamma = 1$ we recover the classical solution related to a Newtonian fluid with time-dependent viscosity and for $\gamma = 0$ the solid behaviour $\epsilon \sim \sigma_0 /\eta_0$.
	
	\begin{figure}[h!]
	\centering
	\includegraphics[scale=0.35]{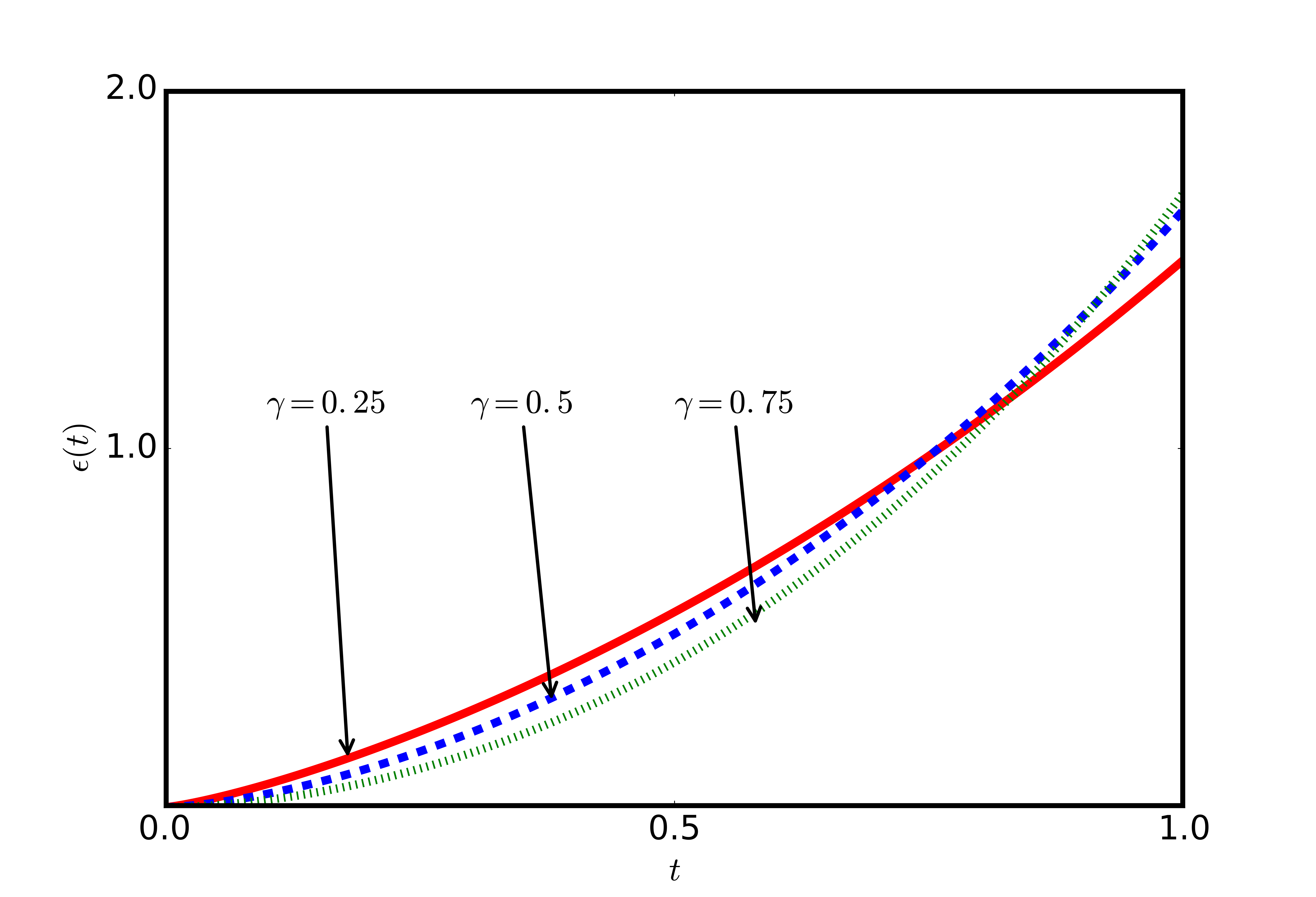}
	\caption{Plot of the strain response to the creep experiment in \eqref{3.8}, setting $\sigma _0$, $\tau$, $\eta _0$ and $\alpha$ to one.}
	 \label{fig-2}
	\end{figure}

\end{example}

\begin{example}[Mittag-Leffler function]
Let us now consider the case where
\be \label{ML}
f(t) = \frac{1}{\alpha} \, E_\nu (\alpha \, t) := \frac{1}{\alpha} \, \sum _{k=0} ^\infty \frac{(\alpha \, t)^k}{\Gamma (\nu \, k + 1)} \, , \quad \nu \in (0, 1] \, ,
\ee
where $\alpha > 0$ is a constant such that $[\alpha] = \mbox{time}^{-1}$.

Clearly, the series representation of the Mittag-Leffler function appearing in \eqref{ML} is uniformly convergent for $t>0$ and $\nu \in (0, 1]$. Besides, if one preforms the derivative term by term of $f(t)$ it is also easy to see that the resulting series is still uniformly convergent for $t>0$ and $\nu \in (0, 1]$, thus one can conclude that
\be 
f'(t) = \frac{1}{\alpha} \, \frac{d}{dt} E _\nu (\alpha \, t) = \sum _{k=0} ^\infty \frac{(k+1) (\alpha \, t)^k}{\Gamma (\nu \, k + \nu + 1)} \, , 
\quad \nu \in (0, 1] \, .
\ee
The latter, being a series with positive terms, implies $f'(t) > 0$ for all $t>0$. Besides, it is also easy to see that $f'(0+) = 1 / \Gamma (\nu + 1)$.

	Then, using the result in \eqref{4} one finds (see \figurename~\ref{fig-3})
	\be \label{3.11}
	\epsilon (t) = \frac{1}{\Gamma (\gamma + 1)} \frac{\sigma _0}{\eta _0 \, \tau ^{\gamma -1}} \, \left[\frac{E_\nu (\alpha \, t) - 1}{\alpha}\right]^{\gamma + 1} \, ,
	\ee
	with $\nu \in (0, 1]$, $\alpha > 0$ and $t>0$.
	
	\begin{figure}[h!]
	\centering
	\includegraphics[scale=0.35]{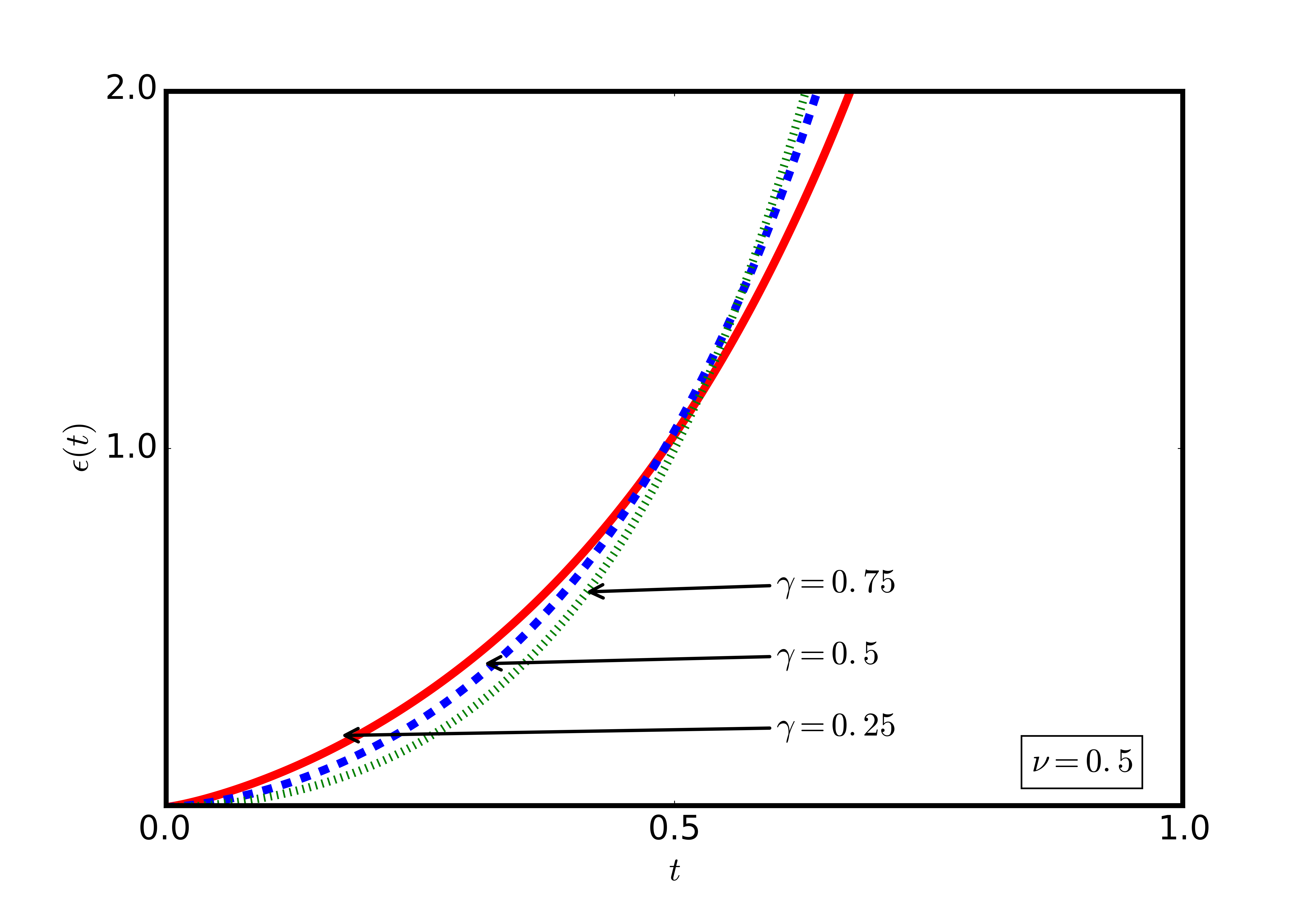}
	\caption{Plot of the strain response to the creep experiment in \eqref{3.11}, setting $\sigma _0$, $\tau$, $\eta _0$ and $\alpha$ to one.}
	 \label{fig-3}
	\end{figure}

\end{example}      

\begin{rmk}
The plots in \figurename~\ref{fig-1} and \ref{fig-3} have been obtained by means of a simplified numerical algorithm based on \cite{Garrappa1,Garrappa2}.
\end{rmk}
    
	\section{Conclusions}
	Nowadays, fluids with time-dependent viscosity play a central role in modern material science and engineering. Indeed, this class of shear-thickening non-Newtonian fluids keep being subjected to a shear stress throughout their whole stress history.
	
	In this paper we have suggested an interesting framework whose aim is to provide a simple mathematical scheme to physically model the emergent global structure of these kind of peculiar material.
	
	In particular, after reviewing the general feature of the fractional derivative of a function with respect to another function, we have provided three different example of constitutive equations with time-dependent viscosity by inducing a minimal modification of the renowned Scott-Blair model of linear viscoelasticity.
	
	Thus considering that both the Newton and the Scott-Blair dashpots represent the fundamental building blocks of any standard model of linear viscoelasticity, we strongly believe that the approach presented here deserves some further studies and experimental investigations, as it might have some relevant consequences for the non-linear theory of viscoelasticity.

\section*{Acknowledgments}	
	The work of the authors has been carried out in the framework of the activities of the National Group for Mathematical Physics (GNFM, IN$\delta$AM).

\end{document}